\documentclass{ifacconf}

\usepackage{graphicx}      
\usepackage{natbib}        
\usepackage{amsmath,amssymb,amsfonts}
\usepackage[table,xcdraw]{xcolor}
\usepackage{pdfpages}
\usepackage{svg}
\usepackage{mathtools}
\usepackage{blkarray, bigstrut}

\usepackage{supertabular,booktabs}
\usepackage{multicol}
\usepackage{float}
\usepackage{multirow}
\usepackage{algorithmic}
\usepackage{graphicx}
\usepackage{textcomp}
\usepackage{diagbox}

\begin{document}
\begin{frontmatter}

\title{A whole-body multi-scale mathematical model for dynamic simulation of the metabolism in man
} 

\thanks[footnoteinfo]{
Corresponding author: J.B. J{\o}rgensen (e-mail: jbjo@dtu.dk). 
}

\author{Peter Emil Carstensen,} 
\author{Jacob Bendsen,} 
\author{Asbjørn Thode Reenberg,}
\author{Tobias K. S. Ritschel,}
\author{John Bagterp Jørgensen}

\address{Department of Applied Mathematics and Computer Science, \\ Technical University of Denmark, DK-2800 Kgs. Lyngby, Denmark}

\begin{abstract}                
We propose a whole-body model of the metabolism in man as well as a generalized approach for modeling metabolic networks. Using this approach, we are able to write a large metabolic network in a systematic and compact way. We demonstrate the approach using a whole-body model of the metabolism of the three macronutrients, carbohydrates, proteins and lipids. The model contains 7 organs, 16 metabolites and 31 enzymatic reactions. All reaction rates are described by Michaelis-Menten kinetics with an addition of a hormonal regulator based on the two hormones insulin and glucagon. We incorporate ingestion of food in order to simulate metabolite concentrations during the feed-fast cycle. The model can simulate several days due to the inclusion of storage forms (glycogen, muscle protein and lipid droplets), that can be depleted if food is not ingested regularly. A physiological model incorporating complex cellular metabolism and whole-body mass dynamics can be used in virtual clinical trials. Such trials can be used to improve the development of medicine, treatment strategies such as control algorithms, and increase the likelihood of a successful clinical trial.
\end{abstract}

\begin{keyword}
Mathematical modeling, metabolism, systems biology, cyber-medical systems, multi-scale modeling, quantitative systems pharmacology

\end{keyword}

\end{frontmatter}

\section{Introduction}
In the human body, metabolites are constantly formed and broken down through a vast number of reactions. The metabolism functions in an organized manner to keep the body alive \citep{miesfeld_mcevoy_2017}. Whole-body modeling uses exactly this idea to describe the human body as a collective unit. Doing so can provide a predicted concentration of a metabolite in any specific organ, which is relevant e.g. to PK/PD drug development \citep{derendorf_schmidt_rowland_tozer_2020}. 

Depending on the modeling objective (i.e. the intended use of the model), there are several ways to model the human metabolism. We look at the system of organs and blood vessels as a whole-body model. The enzymatic reactions in the metabolic network occur in the organs, and the organs are connected with each other through the blood vessels. As the organs are not identical, different reactions and reaction rates are defined based on their role in the metabolism. Following this approach, it is possible to simulate the metabolism of man under various conditions. 

There exist whole-body models in today's literature that describe the metabolism in man at different levels of complexity. \citet{sorensen_1978} refined simple and inadequate models with focus on glucose, insulin and glucagon dynamics, using a simple whole-body model. More complex models have been developed in recent years. \citet{panunzi_pompa_borri_piemonte_gaetano_2020} focused on extending the model proposed by \citet{sorensen_1978}, by adding food intake. Other authors \citep{kim_saidel_cabrera_2006,dash_li_kim_saidel_cabrera_2008, KURATA2021102101} use whole-body models and expand into stoichiometry to include several metabolites. A simple and intuitive approach, in which the mathematical equations can easily be incorporated, is not readily available. By utilizing previous work and modeling principles described by \citet{yasemi_jolicoeur_2021}, we propose such a mathematical approach. 

In this work, we describe a whole-body model of the metabolism in man. Furthermore, we demonstrate a general, systematic and intuitive way to formulate the model equations. The five key organs included are the brain, the heart and lungs, the liver, the gut and the kidney. Further, the muscle tissue and the adipose tissue are each simplified as a single compartment and can therefore be considered as an organ. This results in a total of seven organs. The metabolism inside the organs are explained by the stoichiometry of the enzymatic reactions. We use Michaelis-Menten kinetics to describe the enzymatic reactions. Using the model, we simulate the feed-fast cycle to investigate how prolonged fasting affects selected metabolite concentrations and glucose flux. Further, we simulate intermittent fasting to investigate how it affects the carbohydrate, the protein and the lipid storage.

The remaining part of this paper is structured as follows. Section 2 describes the approach for whole-body modeling. In section 3, a biological model is formulated using the mathematical approach. Section 4 presents the simulation results, and we discuss our formulated model and assumptions in Section 5. Finally, Section 6 concludes on our findings.
\section{Mathematical approach}
\label{sec:mathmaticalmodel}
The general model is described as a system in which metabolites flow in, are metabolized, and flow out. The dynamics of a single compartment is defined by the general differential equation
\begin{equation}
\label{eq:general_eq}
    V \frac{dC}{dt} = M (Q_{in}C_{in} - Q_{out}C) + RV , 
\end{equation}
where $V$ is the volume, $C$ is a vector containing the concentration of the metabolites, $M$ is the external and internal component ordering, $Q_{in}$ is the flow rate of what goes in, $C_{in}$ is a vector containing the concentration of the metabolites that flow in, $Q_{out}$ is the flow rate of what goes out and $R$ is the production rates. The compartments are coupled through concentration gradients in the blood vessels that connect the compartments. $M$ is a square matrix containing only ones and zeros in the diagonal corresponding to the metabolites distributed through the blood vessels (circulating metabolites). For instance, the circulating metabolite, $C_i$, it corresponds to $M_{i,i}=1$. The production rate $R$ is incorporated as a vector defined by
\begin{equation}
    R = (T S)' T r ,
\end{equation}
where $T$ is a matrix of reactions that occur, $S$ is a stoichiometric matrix containing all reactions and $r$ is a vector with the kinetics for the reactions. $T$ contains ones and zeros corresponding to which reactions from the stoichiometric matrix, that are present in the compartment. For instance, a compartment which involves reaction $1,3$ and $5$ from the stoichiometric matrix have $T_{1,1} = 1, T_{2,3} = 1, T_{3,5} = 1$, and zeros elsewhere. The reaction rate vector, $r$, is a function of the concentration of each metabolite:
\begin{equation}
    r = r(C).
\end{equation}
To utilize \eqref{eq:general_eq}, in a whole-body model, it must be formulated for each compartment.

\section{Model}
\label{sec:model}

\begin{figure}[tb]
    \centering
    \includegraphics[width=0.9\columnwidth, keepaspectratio]{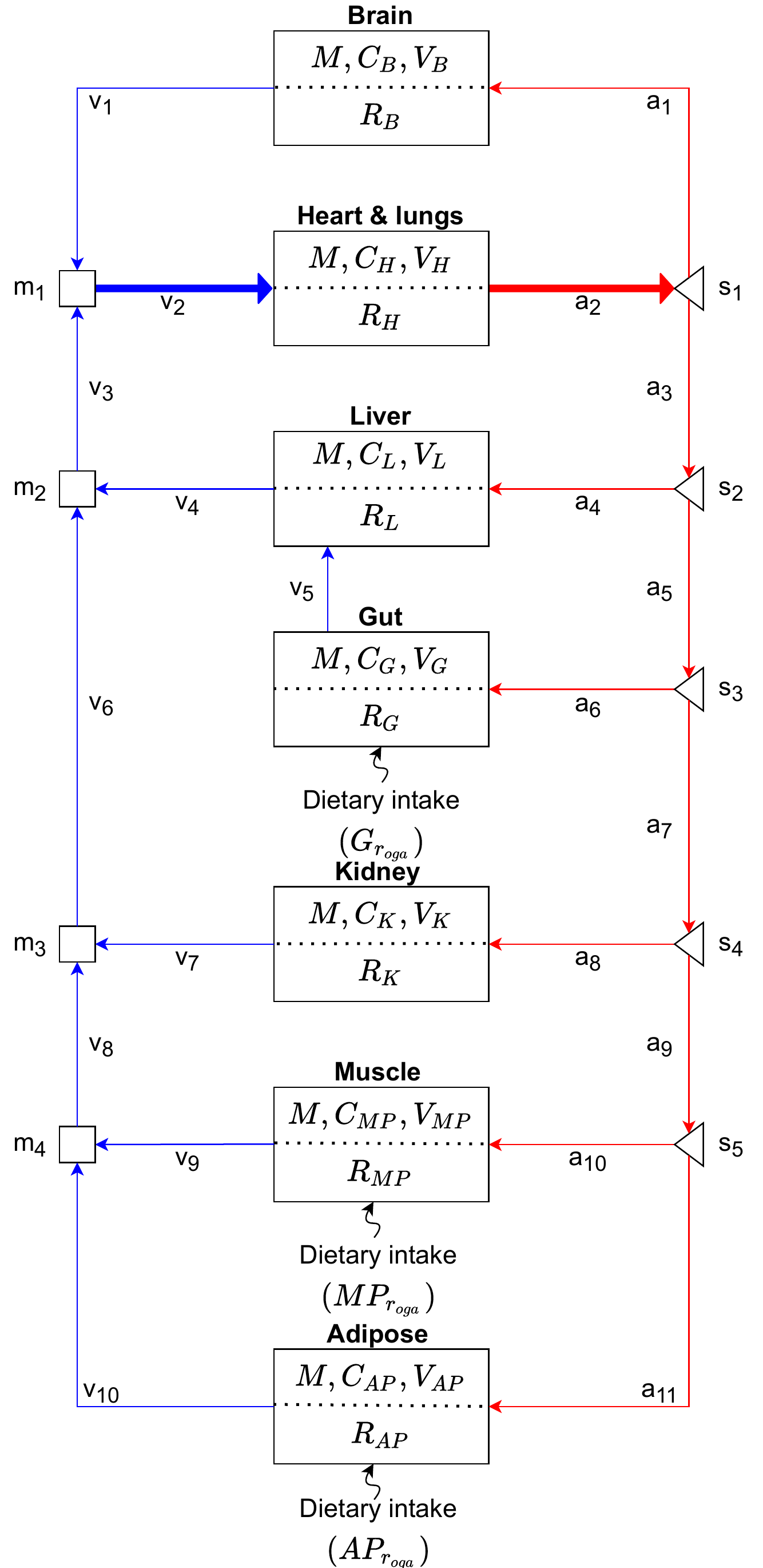}
    \caption{Schematic representation of the whole-body model. Solid arrows represent blood circulation, where the right side is the arteries and the left side the veins.
    Thick arrows, $a_2$ and $v_2$, represents joining of flows from other organs.
    $M$,$C_k$,$V_k$ represents the blood tissue exchange and $R_k$ represents the reactions happening inside the cell. The dotted lines in the compartments suggest free diffusion, as cell-permeability is not included.}
    \label{fig:Bends_cars_flow_diagram}
\end{figure}

We now present a model containing 7 compartments, 16 metabolites and 31 reactions including the hormonal effect from two signal molecules, insulin and glucagon, on specific tissues. Fig. \ref{fig:Bends_cars_flow_diagram} shows a flow diagram of the whole-body model. The blood circulation is split into two parts, the arteries (red) and the veins (blue). The total blood flow, $Q$, is at all times preserved, and the local blood flow can be calculated using mixers and splitters. They are either explicitly modelled as organs, or implicitly as shown by the triangle (splitter) or square (mixer). A splitter divides the blood flow so $a_2 = a_1 + a_3$, and a mixer combines the flood flow so $v_1+v_3 = v_2$. As blood flow is preserved, the flow coming into a compartment is the same as the flow coming out of a compartment. 

\begin{figure}[tb]
    \centering
    \includegraphics[width=1\columnwidth]{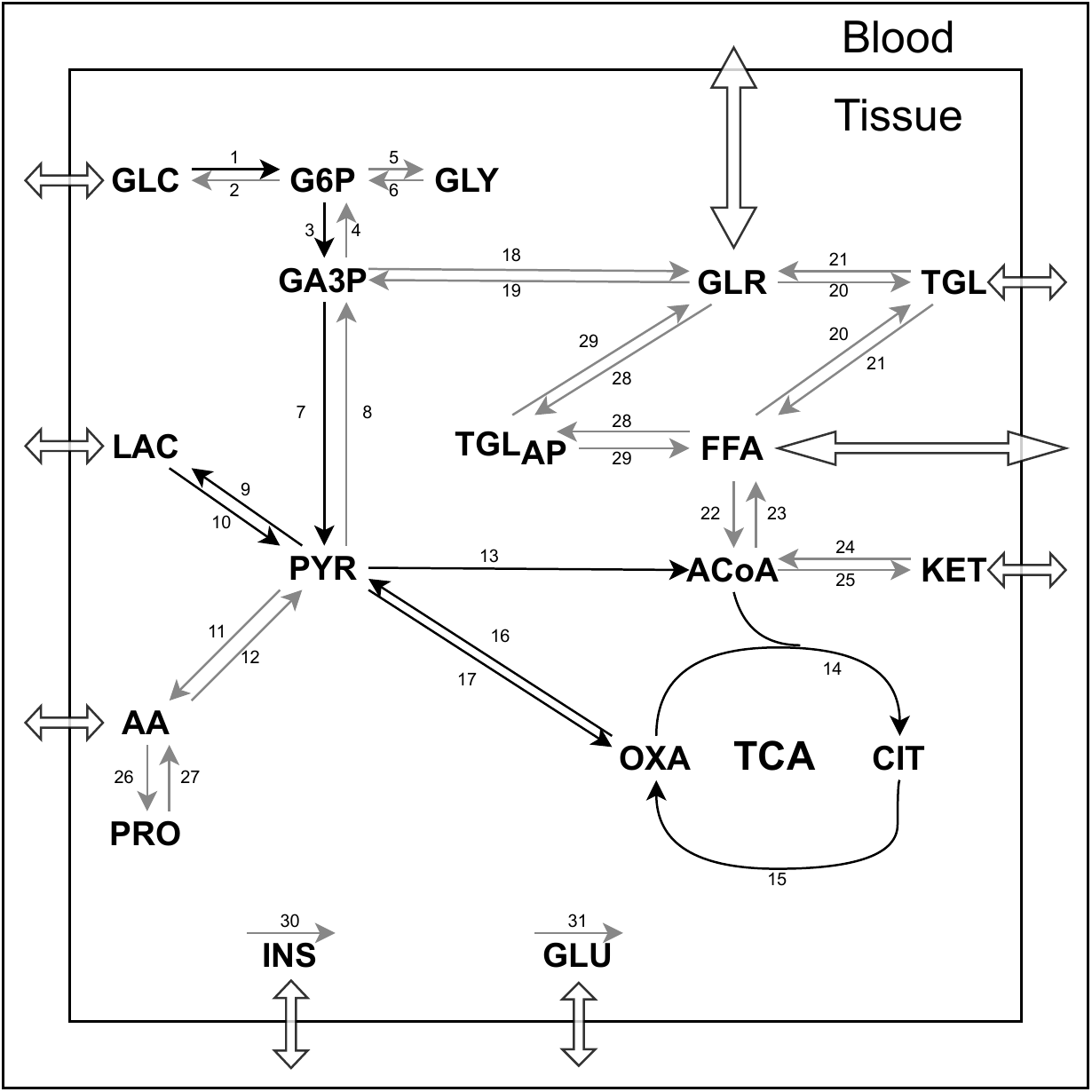}
    \caption{Diagram of metabolic pathways in cell. The hollow double-sided arrows indicate that the metabolite is distributed through the blood. The black arrows indicate reactions that happen in all organs. The grey arrows indicate reactions that only happen in some organs. Metabolites are shown as 2-4 letter abbreviations. The stoichiometric matrix is formed from the numbering of the reactions.}
    \label{fig:metabolic_map}
\end{figure}
We create the model using the same methodology as described in Section \ref{sec:mathmaticalmodel}. The purpose is to describe the energy metabolism of the three macronutrients, carbohydrates, proteins and lipids. We include the major biochemical pathways, shown in Fig. \ref{fig:metabolic_map}, which these macronutrients are a part of, in order to simulate their behaviour under various conditions. Insulin and glucagon are included to add stability and avoid large transient periods in the simulation after food intake, as they are important anabolic and catabolic hormones, respectively. From Fig. \ref{fig:metabolic_map}, it follows that many pathways are cell specific. 10 of 31 reactions occur in all organs. The remaining 21 are tissue specific, as the different organs each have a specialized role. Insulin and glucagon secretion/clearance are incorporated as a single reaction each for simplicity, as they are not created by nor metabolised into any included metabolite. Their reaction rates are inherited from \cite{sorensen_1978}, as described in Section \ref{sec:hormonal_model}.

The most accurate mathematical equations to describe each of these 31 metabolic reactions are not necessarily known, and vary from individual to individual. Therefore each of these reactions have been simplified. In the model, we describe the reaction rates using Michaelis-Menten kinetics. Michaelis-Menten kinetics are especially useful for describing enzyme reactions, as enzyme reactions have an upper limit on the reaction rate. Alternatives to Michaelis-Menten kinetics include altering the mathematical formula in the reaction rate vector to e.g. positive hyperbolic tangent functions as in \cite{sorensen_1978} or simple first-order kinetics from \cite{kim_saidel_cabrera_2006}. If a reaction is excluded from an organ, it does not mean that the reaction never happens in reality. Instead, it means that the reaction is excluded for simplicity. An example is glycogen formation that also happens in the brain, heart and adipose tissue, but it is in such small quantities, that it becomes negligible \citep{gropper_smith_carr_2018}. 

\subsection{Inclusion of a hormonal model}
\label{sec:hormonal_model}

\cite{sorensen_1978} employs a very simple glucagon model that describes the pancreatic glucagon release. Sorensen found that the glucagon release and clearance could be described adequately by a one compartment model. Sorensen also formulated an insulin model described by a six compartment model. While it is possible to include the glucagon model as a single compartment model and insulin as a six compartment model, similar to \cite{panunzi_pompa_borri_piemonte_gaetano_2020} and \cite{sorensen_1978}, we include insulin and glucagon in a whole-body seven compartment model. Insulin and glucagon are therefore included in the stoichiometric matrix as metabolites in order to calculate their production rates. 

\setlength\extrarowheight{3pt}
\begin{table}[tb]
\centering
\caption{Reactions affected by insulin and glucagon \citep{ gropper_smith_carr_2018, miesfeld_mcevoy_2017}. $\Uparrow$ symbolizes increased stimulation and $\Downarrow$ symbolizes decreased stimulation.}
\label{tab:insulin_glucagon_reactions}
\resizebox{\columnwidth}{!}{%

\begin{tabular}{lcccc}
    \textbf{Hormone}  & \textbf{Effect} & \textbf{\# R} & \textbf{Reaction} & \textbf{Affected Organs} \\[0.5ex] \hline
    \textbf{Insulin}  & $\Uparrow$ & 1 & GLC $\rightarrow$ G6P  & Liver, muscle, adipose tissue \\ \cmidrule{3-5}
                    &  $\Uparrow$ & 3 & G6P $\rightarrow$ 2 GA3P  & Liver, muscle tissue \\ \cmidrule{3-5} 
                    &  $\Uparrow$ & 5 & G6P $\rightarrow$ GLY  & Liver, muscle tissue \\ \cmidrule{3-5} 
                    & $\Uparrow$ & 13 &  PYR $\rightarrow$ ACoA  & Liver, muscle tissue \\ \cmidrule{3-5} 
                    & $\Uparrow$ & 21 & TGL $\rightarrow$ 3 FFA + GLR  & Adipose tissue \\ \cmidrule{3-5} 
                    & $\Uparrow$ & 23 & 7 ACoA $\rightarrow$ FFA  & Liver \\ \cmidrule{3-5} 
                    & $\Uparrow$ & 26 & AA $\rightarrow$ PRO   & Muscle tissue \\ \cmidrule{3-5} 
                    & $\Uparrow$ & 28 & 3 FFA + GLR $\rightarrow$ TGL$_\mathrm{AP}$   & Adipose tissue \\ \cmidrule{2-5} 
                    & $\Downarrow$ & 4 & 2 GA3P $\rightarrow$ G6P  & Liver \\ \cmidrule{3-5} 
                    & $\Downarrow$ & 6 & GLY $\rightarrow$ G6P    & Liver, muscle tissue \\ \cmidrule{3-5}
                    & $\Downarrow$ & 27 & PRO $\rightarrow$ AA    & Muscle tissue           \\ \cmidrule{3-5}
                    & $\Downarrow$ & 29 & TGL$_\mathrm{AP}$ $\rightarrow$ 3 FFA + GLR     & Adipose tissue \\ 
    \midrule 
    \textbf{Glucagon} & $\Uparrow$ & 4 &  2 GA3P $\rightarrow$ G6P & Liver \\ \cmidrule{3-5} 
                    & $\Uparrow$ & 6 & GLY $\rightarrow$ G6P & Liver \\ \cmidrule{3-5} 
                    & $\Uparrow$ & 29 &  TGL$_\mathrm{AP}$ $\rightarrow$ 3 FFA + GLR & Adipose tissue \\ \cmidrule{2-5} 
                    & $\Downarrow$ & 3 & G6P $\rightarrow$ 2 GA3P & Liver \\ \cmidrule{3-5}
                    & $\Downarrow$ & 5 & G6P $\rightarrow$ GLY & Liver \\  
   \bottomrule
  \end{tabular}%
}
\end{table}
\setlength\extrarowheight{0pt}
Table \ref{tab:insulin_glucagon_reactions} contains the reactions that are affected by insulin and glucagon, as well as which organ is affected. While the qualitative effect of insulin and glucagon is known (see Table \ref{tab:insulin_glucagon_reactions}), the related model parameters for insulin and glucagon are not necessarily known. \cite{miesfeld_mcevoy_2017} describe qualitatively which enzymes and hence reactions that insulin and glucagon affect. Accordingly, given the reciprocal effects of insulin ($I$) and glucagon ($\Gamma$), we use a simple function where the ratio between them allows us to determine whether or not they have an effect on the system. These functions alter the production rates of the reactions in Table \ref{tab:insulin_glucagon_reactions}. If only insulin have an effect on the reaction, the functions are
\begin{equation}
    \textbf{Insulin activation:} \left(\frac{I_k}{I_k^B}\right)^{\mu_j} V_{max_j} ,
\end{equation}
\begin{equation}
    \textbf{Insulin inhibition:} \left(\frac{I_k^B}{I_k}\right)^{\mu_j} V_{max_j} ,
\end{equation}
and if both insulin and glucagon have an effect on the same reaction (though with opposite effects), the insulin-to-glucagon or glucagon-to-insulin ratio is used instead,
\begin{equation}
    \textbf{Insulin-to-Glucagon stimulation:} \left(\frac{\Gamma_k^B}{\Gamma_k} \hspace{1mm} \frac{I_k}{I_k^B}\right)^{\mu_j} V_{max_j} ,
\end{equation}
\begin{equation}
    \textbf{Glucagon-to-insulin stimulation:} \left(\frac{\Gamma_k}{\Gamma_k^B} \hspace{1mm} \frac{I_k^B}{I_k}\right)^{\mu_j} V_{max_j} ,
\end{equation}
where $j$ is the reaction, e.g. \textit{GLC$\rightarrow$G6P}, $k$ is the compartment and the superscript $B$ indicates that it is the basal-value. $V_{max_j}$ is the maximum rate in the Michaelis-Menten kinetics and $mu_j$ is a scaling parameter for the hormonal effect. These simple functions are equal to $V_{max_j}$ at steady state, which occurs when the blood glucose concentration is at $5\,\mathrm{mmol}/\mathrm{L}$, such that it is independent of $\mu_j$ in steady-state. This is an important property, since many of the parameters used in the reactions are estimated based on metabolite homeostasis. 
\subsection{Inclusion of a modified SIMO-model}  

\begin{figure}[tb]
    \centering
    \includegraphics[width=0.9\columnwidth]{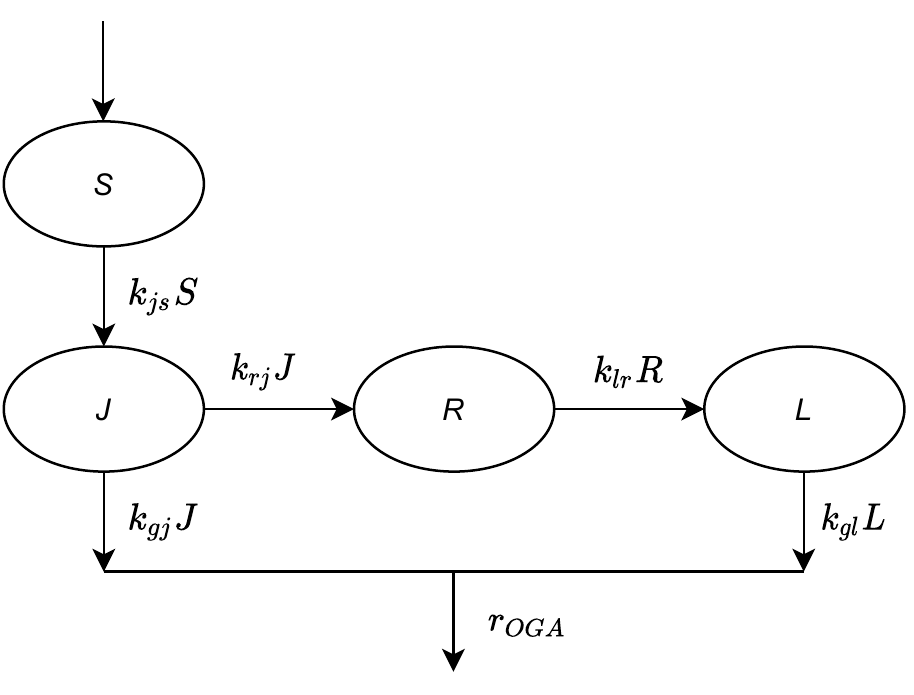}
    \caption{Schematic representation of the digestive tract. \textit{S} represents the amount of macronutrients in the stomach, \textit{J} the jejunum, \textit{R} a delay and \textit{L} the amount in the ileum. $r_{OGA}$ is a vector describing the uptake of the three macronutrients.}
    \label{fig:simo}
\end{figure}

A modified version of the simple interdependent glucose/insulin model, SIMO \citep{panunzi_pompa_borri_piemonte_gaetano_2020}, is used to include the uptake of glucose, amino acids and lipids from ones diet. The rates are the same for all macronutrients, which is not physiologically accurate. It is known from \cite{miesfeld_mcevoy_2017} that the macronutrients differ in terms of absorption given their molecular differences. Due to lack of data or other mathematical models describing the uptake of macronutrients, we adopt the SIMO model as it is the simplest choice. The resulting uptake of macronutrients is represented by
\begin{equation}
    r_{OGA} = k_{gj}J + k_{gl}L = \begin{bmatrix}
           k_{gj}J_{GLC} + k_{gl}L_{GLC} \\
           k_{gj}J_{AA} + k_{gl}L_{AA} \\
           k_{gj}J_{TGL} + k_{gl}L_{TGL}
         \end{bmatrix},
\end{equation}
where $k_{gj}$ and $k_{gl}$ are uptake rates and $r_{OGA}$ is a 3$\times$1 vector, where the first two macronutrients, i.e. glucose and amino acids, are taken up by the gut. This is not the case for triglycerides, which enter the lymphatic system as chylomicrons and is transported to muscle and adipose tissue before it enters the blood circulation \citep{gropper_smith_carr_2018}. The triglycerides ($TGL$) from $r_{OGA}$ is then delivered to muscle and adipose tissue, where a 50/50\% distribution is assumed in the two tissues. As the specific uptake rates of amino acids and lipids are not explicitly known from the SIMO model, we utilize the uptake rate $k_{gj}$ and $k_{gl}$ from glucose. It is included in the model as the parameters:
\begin{align*}
    \text{Gut: } &  G_{r_{OGA}} \\
    \text{Muscle: } & MP_{r_{OGA}} \\
    \text{Adipose: } & AP_{r_{OGA}}
\end{align*}
The modified SIMO model is included as additional inputs in the differential equations and not directly into the production rate vector $R$, as the modified SIMO model provides external inputs from meal consumption and is seen as an extension to the general methodology. 

\section{Simulation results}
\label{sec:simulationresults}

\begin{figure}[tb]
    \centering
    \includegraphics[width=\columnwidth]{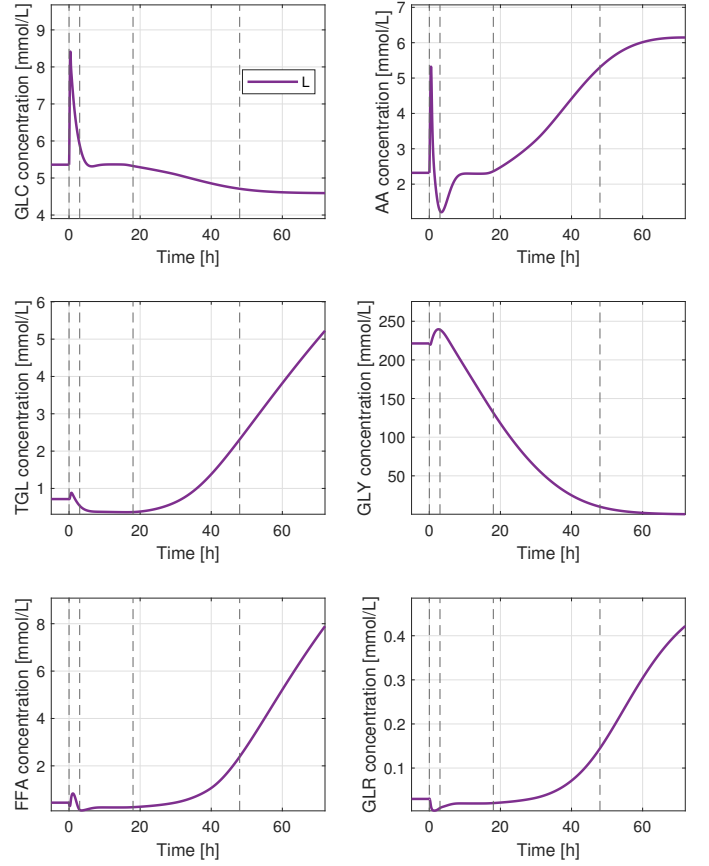}
    \caption{The metabolite concentrations of glucose ($GLC$), amino acids ($AA$), triglycerides ($TGL$), glycogen storage ($GLY$), free fatty acids ($FFA$) and glycerol ($GLR$) in the liver after an initial meal of 60 g glucose, 24 g protein and 16 g fat and an accompanying fasting for 72 hours.}
    \label{fig:Liver_metabolites}
\end{figure}

We now simulate the model presented in Section \ref{sec:model} in MATLAB. Initially we present the model with a single meal ingested at steady-state, and follow the development of the metabolites as an in silico patient refrains from eating or doing physical activity for the next 72 hours. Fig. \ref{fig:Liver_metabolites} shows the glucose concentration, the amino acids, the triglycerides, the glycerol, the free fatty acids and the glycogen concentration. The plots for $GLC$, $AA$ and $TGL$ show the metabolites that are ingested through the modified SIMO model. After the initial meal, we see that, the glucose concentration initially rises and then returns to the baseline, where it remains constant for roughly 10 hours. The glucose concentration then starts to decrease as glycogen storage is reduced substantially. While the glucose concentration diminishes, the concentrations of the other metabolites start increasing. This is especially the glycerol concentration, the free fatty acids and the triglycerides, as the patient enters the starvation stage, where lipids are the main energy source for several organs. An increase in the triglycerides is seen during starvation as the triglyceride storage in the adipose tissue, $TGL_{AP}$, reduces. The free fatty acid levels increase substantially during starvation, which is to be expected based on articles that investigate the effects of starvation \citep{unger_eisentraut_madison_1963, yaffe_1980A}.
\begin{figure}[tb]
    \centering
    \includegraphics[width=0.9\columnwidth]{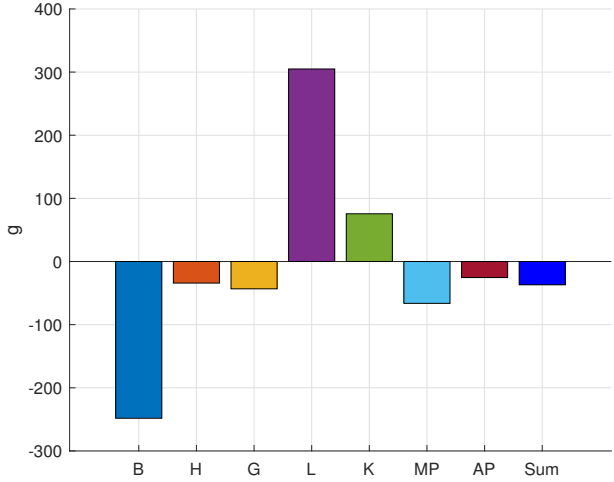}
    \caption{Sum of the glucose fluxes in every organ after an initial meal and accompanying fasting for 72 hours. The organs are the brain (B), the heart and lungs (H), the gut (G), the liver (L), the kidneys (K), the muscle tissue (MP) and the adipose tissue (AP). The total sum of all the glucose fluxes is the rightmost column denoted 'Sum'.}
    \label{fig:glucose_fluxa}
\end{figure}

\begin{figure}[tb]
    \centering
    \includegraphics[width=0.9\columnwidth]{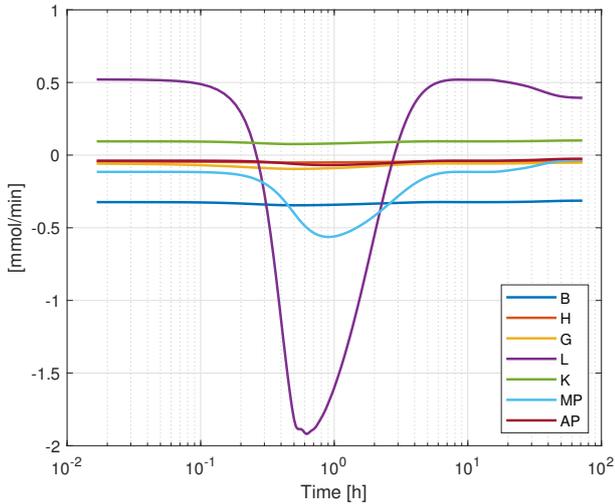}
    \caption[short]{The glucose fluxes in every organ after an initial meal and accompanying fasting for 72 hours. It shows the fluxes over time with a logarithmic scale on the x-axis.}
    \label{fig:glucose_fluxb}
\end{figure}
\subsection{Fluxes inside the cells}
While it is important to keep in mind that these simulation results are not physiologically accurate for all metabolites, we can plot the fluxes in order to investigate the dynamics. Fig. \ref{fig:glucose_fluxa} shows that the brain is a major consumer of glucose, and the liver is a major exporter of glucose, which is also what is expected from \cite{miesfeld_mcevoy_2017}. The rightmost column, Sum, shows the net flux of glucose in all organs, resulting in an overall consumption corresponding to a meal containing 60 g glucose. Fig. \ref{fig:glucose_fluxb} shows which organs are the consumers of glucose over time. There is a large drop initially for the glucose flux in the liver and the muscle tissue as food is ingested and the blood glucose concentration is high. This is especially seen in the insulin stimulated tissues, as the effect described in Table \ref{tab:insulin_glucagon_reactions} shows an increase in the glucose uptake as a result of high concentrations of insulin. As the blood glucose concentration reduces, the liver again produces glucose for the other organs in order to keep homeostasis, and thus increases its flux to a positive value.

\subsection{Simulation of intermittent fasting}
\begin{figure}[tb]
    \centering
    \includegraphics[width=\columnwidth]{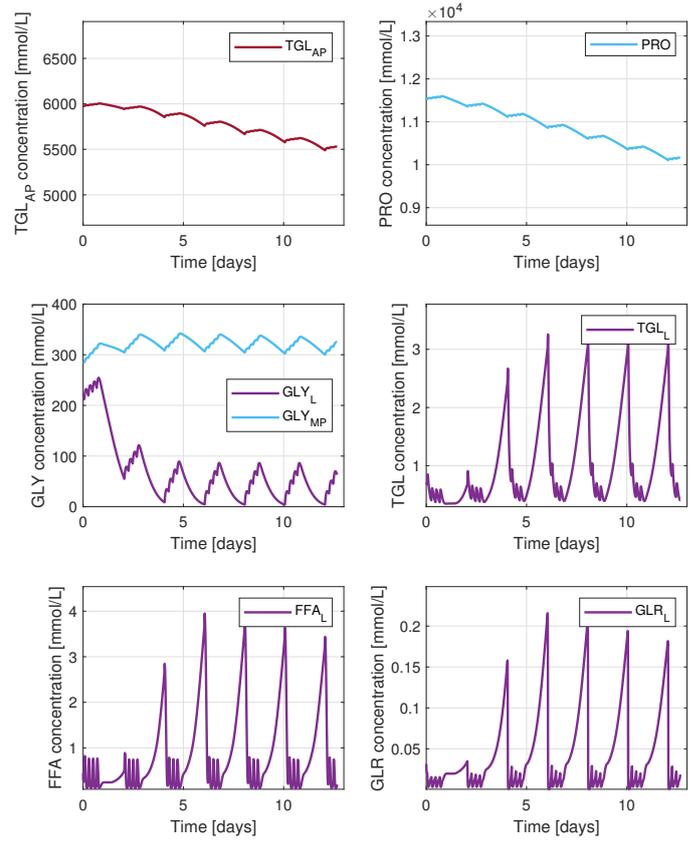}
    \caption{The metabolite concentrations of the fat storage in the adipose tissue ($TGL_{AP}$), the protein storage in the muscle tissue ($PRO$), the glycogen storage in the liver ($GLY_L$) and in the muscle tissue ($GLY_{MP}$), the triglycerides in the liver ($TGL_L$), the free fatty acids in the liver ($FFA_L$) and the glycerol in the liver ($GLR_L$). The simulation is run for 13 days with intermittent fasting every other day.}
    \label{fig:intermittent_fasting}
\end{figure}
Fig. \ref{fig:intermittent_fasting} shows a simulation over 13 days, with regular food intake every other day, resulting in intermittent fasting for 33 hours. We see a decrease in the concentration of lipid droplets, as they are metabolized to provide energy for the body. This is due to the energy balance of the system becoming skewed from a roughly equal calory intake/consumption, to a 50\% reduction in calory intake. Further, we see that the protein storage in the muscle tissue is also broken down, as food becomes a relatively scarce source for the body. Within a day we see a large drop in the concentration of glycogen after the last meal. Glycogen is initially used as an energy storage, but gets depleted quickly in the liver. The glycogen in the muscle tissue is relatively stable, as the in silico patient is at rest for the entire simulation. The lipid droplets are converted into $FFA$ and $GLR$ and utilized in different compartments. We see large spikes when the body enters the starvation phase after 18 hours without any food (see also Fig. \ref{fig:Liver_metabolites}). The simulation shows, that following this diet results in an effective weight loss, as $TGL_{AP}$ (body fat) levels decrease. However there are adverse effects to this diet, as muscle proteins are also diminished. By simulating intermittent fasting, we are able to show both the effects of regular meal intake and fasting.

\section{Discussion}
Diffusion across the cell membrane is assumed to happen infinitely fast, as the time it takes for the cell to be in equilibrium with the blood vessels happens at a much faster timescale compared to the timescale of our model. This assumption makes it possible to include organs as one compartment. If transporters and diffusion were to be incorporated in the model, it would be expected to divide the compartments in two, intracellular and extracellular, such that transporters could be explicitly modeled. The concentration is assumed to be the same across the entire compartment. We model each organ as a well-stirred tank. Therefore every metabolite is uniformly distributed in the organs. It is known, that the enzymatic reactions in each cell and organ happen at a fast timescale of $10^{-3} \text{ to } 10^0$ seconds \citep{yasemi_jolicoeur_2021}. We are interested in simulating the model for up to several days. Therefore, we assume that the enzymatic reactions happen at a uniform rate across the spatial organs, allowing a simplification of enzymatic reactions to Michaelis-Menten kinetics.

It could be theorized, given the introduction of more metabolic inputs, that a model like this could be used to gain a better understanding of the dynamical changes in metabolite concentrations of an individual. Both a generalized model could be used, or one tailored to a specific person's parameters, thus introducing personalized modeling behaviour. Simulation of enzymatic defects to represent various diseases, could be done with the current format of the model. However, it would require validation of the model.

\section{Conclusion}
\label{sec:conclusion}

Using a systematic approach for modeling metabolic networks, we have developed a model that is capable of simulating the complex human metabolism. The approach makes it simple to expand the system through changes in the stoichiometric matrix, that match the chemistry and modeling objective. The core of the model is simply which parameters and reaction kinetics that are used in the production rate vector $R_k$. This is illustrated in the model, which involves 16 metabolites and 2 hormones in 7 organs. It results in a total of 126 differential equations. We write these as 7 differential equations, one for each organ, as the entire reaction network is included in $R_k$. As shown, the modeling approach can readily be expanded to incorporate large networks of metabolic reactions in the body. The approach is applicable to not just a physiological model, but can easily be made to fit chemical systems, with containers, tubes and chemical reactions taking place. The modeling objective was to create the basis for a physiological whole-body model that can incorporate cellular metabolic processes, so that qualitative knowledge can be utilized in a quantitative manner and, when adequately tested, used for in silico trials.

\bibliography{ifacconf}

\end{document}